\documentclass[%
 reprint,
superscriptaddress,
 amsmath,amssymb,
 aps,
]{revtex4-2}
\usepackage{ulem} %for \sout
\usepackage{graphicx}% Include figure files
\usepackage{dcolumn}% Align table columns on decimal point
\usepackage{bm}% bold math
\usepackage{xcolor}
\usepackage{gensymb}
\usepackage[detect-all]{siunitx}
\begin{document}

\preprint{APS/123-QED}

\title{Structural properties of epitaxial $\alpha$-U thin films on Ti, Zr, W and Nb}% Force line breaks with \\

\author{R. Nicholls} \email{beckie.nicholls@bristol.ac.uk} \affiliation{H. H. Wills Physics Laboratory, University of Bristol, Bristol BS8 1TL, United Kingdom}%

\author{D. A. Chaney} %\email{daniel.chaney@esrf.fr} 
\affiliation{European Synchrotron Radiation Facility, 71 avenue des Martyrs, 38000 Grenoble, France
}

\author{G. H. Lander} 
\affiliation{H. H. Wills Physics Laboratory, University of Bristol, Bristol BS8 1TL, United Kingdom}%

\author{R. Springell} \affiliation{H. H. Wills Physics Laboratory, University of Bristol, Bristol BS8 1TL, United Kingdom}%
\author{C. Bell}%
 %\email{cb13399@bristol.ac.uk}
\affiliation{H. H. Wills Physics Laboratory, University of Bristol, Bristol BS8 1TL, United Kingdom}%
\date{\today}% It is always \today, today,

\begin{abstract}
Thin layers of orthorhombic uranium ($\alpha$-U) have been grown onto buffered sapphire substrates by d$.$c$.$ magnetron sputtering, resulting in the discovery of new epitaxial matches to Ti(00.1) and Zr(00.1) surfaces. These systems have been characterised by X-ray diffraction and reflectivity and the optimal deposition temperatures have been determined. More advanced structural characterisation of the known Nb(110) and W(110) buffered $\alpha$-U systems has also been carried out, showing that past reports of the domain structures of the U layers are incomplete. The ability of this low symmetry structure to form crystalline matches across a range of crystallographic templates highlights the complexity of U metal epitaxy and points naturally toward studies of the low temperature electronic properties of $\alpha$-U as a function of epitaxial strain.

%\begin{description}
%\end{description}
\end{abstract}

\maketitle

%\tableofcontents

\section{\label{sec:level1}Introduction}
%In the nuclear industry, thin films of actinide materials that are epitaxially locked to a suitable substrate provide a microcosm of fuel/reactor environments, with the added benefit of reduced radioactivity..... Thin films permit the study of surface reactions and kinetics via removal of the `bulk' signal which would dominate in certain synchrotron measurements (\textbf{trying to summarise Gerry's points from his talk about UO2 films}). 

 %, The rapid oxidation of uranium in air and the increased activity of larger samples adds further complexity to the growth and storage of bulk crystals. 
Many of the actinides and their compounds exhibit fascinating condensed matter physics, including a plethora of unusual structural and electronic ground states (e$.$g$.$ complex polymorphism, unconventional magnetic ordering and heavy fermion superconductivity). These properties often arise as a result of the outer-shell 5$f$ electrons being situated on the boundary between itinerancy and localisation \cite{Elgazzar2009HiddenSymmetrybreaking, Hayes2021MulticomponentUTe2, Stewart2017UnconventionalSuperconductivity,Joynt2002TheUPt3, Clark2019PlutoniumHandbook}. The mid-series actinide metals (U, Np, Pu) exemplify these characteristics, with each of the three distinct crystallographic structures adopted by bulk U exhibiting notable collective electronic phenomena at low temperatures \cite{Moore2009NatureMetals,Lander1994TheReview}. 

The thermodynamically stable phase of uranium under ambient conditions is orthorhombic $\alpha$-U ($Cmcm$). This phase is unique amongst the elements, both for its low symmetry crystal structure and for its unusual electronic properties, with bulk $\alpha$-U crystals undergoing a series of three charge density wave (CDW) transitions before entering a superconducting (SC) state. Although the superconducting transition temperature ($T_\textrm{c}$) appears to vary unpredictably with sample crystallinity and purity ($T_\mathrm{c}=0.02-0.78$ K), it is generally accepted that isotropic compressive pressure can be used to suppress the CDW transitions and enhance $T_{\mathrm{c}}$ to a maximum of 2 K near 1.5 GPa \cite{Fowler1967Positive-uranium, OBrien2002Magnetic-uranium, Schmiedeshoff2004Magnetotransport-uranium}. The exact nature of the interaction between the SC and CDW states in $\alpha$-U is yet to be understood, but a combination of bulk and thin film studies have since confirmed that pressure-induced changes to the low temperature states are related primarily to the length of the $a$-axis \cite{Raymond2011UnderstandingCoupling, Springell2014MalleabilityFilms}. 

Epitaxial strain engineering can often be used to explore regions of phase space that are inaccessible in bulk experiments involving uranium. For example, epitaxial layers of $\alpha$-U with $a_\textrm{film}\approx a_\textrm{bulk}$ host an incommensurate `bulk-like' CDW below 43 K, while U layers with $a_\textrm{film}>a_\textrm{bulk}$ (i.e. a strain that would be difficult to attain in bulk crystals) host a near-commensurate CDW with an increased onset temperature of 120 K \cite{Springell2008ElementalFilms,Springell2014MalleabilityFilms}. Compression of the $c$-axis and expansion of the $b$-axis are both predicted to stabilise the CDW state in $\alpha$-U \cite{Xie2021}, but the influence of uniaxial strain along these axes has not yet been explored. %The CDW instability is linked to the softening of the $\Sigma_4$ phonon mode near the $[\frac{1}{2}, 0, 0]$ point in the reciprocal lattice \cite{Raymond2011UnderstandingCoupling}. that may be useful in disentangling these interplay between these two states.

\begin{table*}[t]
\begin{ruledtabular}
\begin{tabular}{cccccccc}
Substrate   &   Buffer   & $T_\textrm{B}$ ($^\circ$C) & $t_\textrm{B}$ ({\AA}) & $T_\textrm{U}$ ($^\circ$C)          & $t_\textrm{U}$ ({\AA}) &     Cap       & $t_\textrm{C}$ ({\AA})   \\
\hline
Al$_2$O$_3$(11.0) & Nb(110)  & 800   &  200   &    600  &  5000 & Nb   & 130     \\
Al$_2$O$_3$(11.0) & W(110)  & 750      & 85    &  450     &     1000     & W         &      90      \\
Al$_2$O$_3$(00.1) & Zr(00.1) & 700      & 220         & 20, 250, 400, 500     &    520      & Nb         &  150          \\
Al$_2$O$_3$(00.1) & Ti(00.1) & 600      & 180     & 20, 200, 400, 600 &     600     & Ir or Ti   & 85 or 180           
\end{tabular}%
\caption{\label{tab:growth_params}Growth temperatures ($T$) and nominal layer thicknesses ($t$) for $\alpha$-U thin films (subscript `U') deposited onto various buffer layers (subscript `B') and substrates. Layers grown without any intentional substrate heating are denoted as 20 $^\circ$C. Each sample has been capped with a passivating layer (thickness $t_\textrm{C}$), deposited at room temperature. Typical room temperature deposition rates for each element were 0.38 (Ti), 0.70 (Zr) 0.35 (Nb), 0.48 (Ir), 0.44 (W) and 1.2-1.4 {\AA}/s (U).}
\end{ruledtabular}
\end{table*}

%\textbf{Remove this paragraph?} The hexagonal close-packed Ti(00.1) buffer layers were initially expected to promote the growth of the metastable \textit{hcp}-U structure ($a_U=2.962$ {\AA} vs. $a_{Ti}=2.95$ {\AA}) reported in Ref.~\cite{Springell2014MalleabilityFilms}. However, highly crystalline layers of the $\alpha$-U(110) phase were instead stabilised with a $c$-axis that is 0.8\% larger than bulk.  We then study Zr(00.1)-buffered systems and find two distinct, temperature dependent epitaxial matches between Zr and $\alpha$-U \textbf{(mention something interesting? strains? intermixing?)}. Synchrotron diffraction studies of the Nb/U system - assumed to be a single domain thin film hosting a 3D bulk-like CDW ($T_{\mathrm{CDW}}=43$ K) - have found evidence for a second structural domain. \textbf{Mention tungsten systems? reword} 

It is also known that deposition onto other crystallographic templates can produce well-ordered overlayers that are difficult to stabilise in the bulk. Recently, crystalline layers of the tetragonal $\beta$-U phase have been stabilised at room temperature via deposition onto Si(111) \cite{Yang2021MicrostructureFilms} and single crystal layers of a pseudo body-centred cubic $\gamma$-U structure have been realised by the co-deposition of U and Mo onto Nb(110) \cite{Chaney2021}. Uranium may also form a `hexagonal close-packed' (hcp) structure that is not found in the bulk when deposited on W(110) \cite{Molodtsov1998DispersionMetal, Boysen1998DispersionCeRh3, Chen2019DirectW110}, Gd(00.1) \cite{Springell2008ElementalFilms,Springell2008PolarizationMultilayers} and Cu(111) or Ir(111) buffer layers, although the U layers in the final two systems gradually transition back into $\alpha$-U \cite{Nicholls2022StructureFilms}.

Given the rich array of nearly degenerate structural ground states, it is often difficult to predict the phase and orientation that a uranium layer will form under specific growth conditions. A key task in this area is, therefore, to examine a range of metallic buffer layers that can be used to stabilise high quality epitaxial layers of each U allotrope. The range of strains, structures and orientations will allow further exploration of their intriguing electronic properties, provided the complex crystallographic domain structures are also fully characterised. In this work we investigate the epitaxy of $\alpha$-U onto two new buffer layers (Ti, Zr) and revisit the Nb and W systems from Ref.~\cite{Ward2008TheUranium} to add new information to the previously reported domain structures. 

Section II of this paper describes the growth and characterisation procedures for each thin film system. Section III explores the structure and orientation of crystalline $\alpha$-U grown onto Ti, Zr and W buffers as discerned from laboratory-based X-ray diffraction (XRD) and X-ray reflectivity (XRR). The epitaxy and interface quality in the Ti/U and Zr/U systems are explored as a function of temperature using these techniques. Also included in Section III are synchrotron X-ray diffraction measurements of epitaxial Nb/$\alpha$-U(110) systems which reveal a which reveal a previous unreported domain. The physical origin of the domain is discussed.

\section{Experimental Methods }

\subsection{Growth of epitaxial $\alpha\mbox{-U}$ films}
All samples in this study were grown using the actinide d$.$c$.$ magnetron sputtering system at the University of Bristol, UK. This ultra-high vacuum system operates at base pressures of the order $10^{-10}$ mbar and contains four sputtering guns inside a load-locked chamber \cite{Springell2023Review}. Substrates are loaded onto an adjustable height stage adjacent to a resistive heater capable of achieving temperatures of up to 850 $^\circ$C. The substrates for epitaxial Ti(00.1) and Zr(00.1) growth were $c$-plane Al$_2$O$_3$(00.1) and the substrates for Nb(110) and W(110) growth were $a$-plane Al$_2$O$_3$(11.0). All substrates (sourced from MTI Corp) were polished to optical grade.

The nominal layer thicknesses and growth temperatures are given in Table~\ref{tab:growth_params}. The buffer growth temperature is denoted as $T_\textrm{B}$ and the uranium growth temperature as $T_\textrm{U}$ with respective film thicknesses, $t_\textrm{B}$ and $t_\textrm{U}$. Each sample was capped with a layer of a corrosion resistant metal deposited at room temperature in order to protect the U from ex-situ oxidation. All layers were deposited using approximately $7.5\times$10$^{-3}$ mbar high purity argon gas as the sputtering medium.

\subsection{Structural characterisation}
Structural characterisation of the Ti/U, Zr/U and W/U systems was performed using a Philips X'Pert diffractometer with a Cu-K$_\alpha$ source. XRR profiles were modelled using the GenX package, where the error bars on each fitting parameter are calculated from a 5\% change in the optimal figure of merit \cite{Bjorck2007GenX:Evolution}. Characterisation of the Nb/U system was performed using the diffuse scattering diffractometer at the ID28 beamline (ESRF, France) \cite{ID28_SS}. Synchrotron data were treated using the CrysAlis Pro software package \cite{Crysalis}, high resolution reciprocal space maps were produced using in-house programs and visualised in the DECTRIS Albula package \cite{ALBULA}. All X-ray measurements were conducted at room temperature.

\section{Results and discussion}

\subsection{Titanium buffered system}
\begin{figure*}[t]
\centering
	\includegraphics[width=1\textwidth]{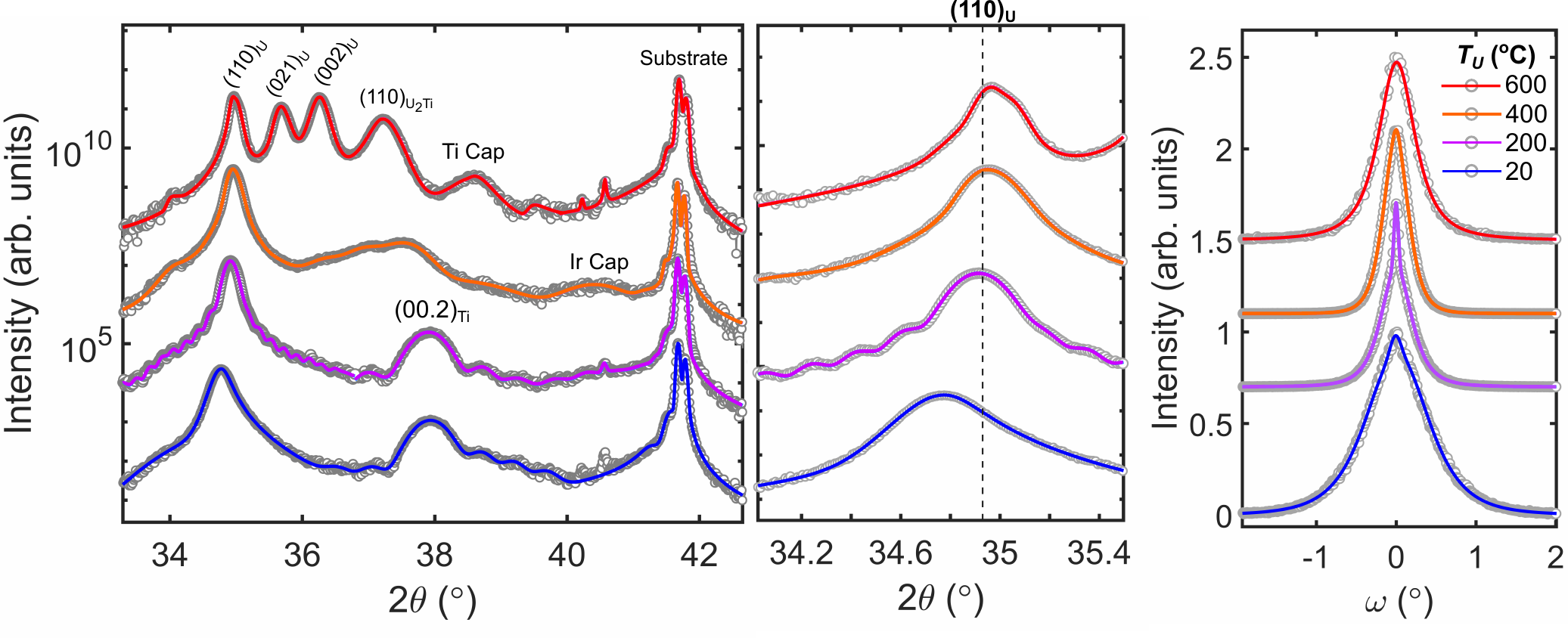}
	\caption{Left: Specular XRD ($2\theta$-$\omega$) scans of layers of U deposited onto Ti(00.1) buffers at various temperatures shown on a logarithmic scale to highlight the Laue fringes. The system transitions from pure, strained $\alpha$-U(110) into a mixture of elemental metals and intermetallic compounds as $T_\textrm{U}$ increases. Center: Detail around $2\theta = 35^{\circ}$ showing a systematic shift of the U(110) peak position with $T_\mathrm{U}$. Vertical dashed line shows the bulk 2$\theta$ position. Right: Evolution of the $\alpha$-U(110) rocking curve. In all cases, datasets are offset vertically for clarity and solid lines are fits to the data.}
	\label{fig:Ti_HAs_all}
\end{figure*}

Titanium was sputtered onto Al$_2$O$_3$(00.1) substrates at 600~$^\circ$C to produce epitaxial hexagonal close-packed Ti(00.1) layers with the in-plane relationship $[10.0]_{\mathrm{Ti}}\parallel [1\bar{1}.0]_{\mathrm{Al}_2\mathrm{O}_3}$ and thickness 180 {\AA}. Uranium layers with a nominal thickness of $520$ {\AA} were subsequently deposited at various temperatures. Fig.~\ref{fig:Ti_HAs_all} shows the coupled $2\theta$-$\omega$ scans and rocking curves from the temperature series. XRR profiles with discernible Kiessig fringes are included as Supplemental Information \cite{SupplementalInformation}. Table \ref{table:fitparams} summarises the $d_{110}$ spacings, widths of each rocking curve ($\Delta\omega$) and the XRR-derived root-mean-square roughnesses ($\sigma$) across the series.

\begin{table}[b]
\begin{ruledtabular}
\begin{tabular}{ccccc} 
$T_\textrm{U}$ ($^{\circ}$C)  & $d_{110}$ ({\AA}) & $\Delta\omega_\mathrm{U}$ ($^\circ$) & $V_\mathrm{XRD}$ ({\AA}$^3$)  & $\sigma_{\mathrm{U}}$ ({\AA})  \\  \hline
  20   & 2.578 & \uline{0.933}  & 21.207  & $16\pm3$  \\
200   & 2.568 & \uline{0.455} & 20.885   & $14\pm3$  \\ 
400   & 2.565 & 0.370  & - & -   \\ 
600   & 2.564 & 0.543 &  - & -    \\
\end{tabular}
\caption{\label{table:fitparams}Parameters extracted from XRD/XRR measurements for the Ti/U temperature series. Values are not provided in cases where U-Ti intermixing has destroyed the coherent Ti buffer. The rocking curve width ($\Delta\omega$) represents the full-width at half maximum (FWHM) of the broadest component in the peak profile; samples \uline{underlined} also contain a resolution-limited Gaussian (see main text for details). Root-mean square roughness ($\sigma$) values were extracted from GenX models fitted to XRR data.  }
\end{ruledtabular}
\end{table}

At the two lowest tested deposition temperatures, the U layer forms crystalline $\alpha$-U(110). At $T_\textrm{U}=20$ $^\circ$C, the out-of-plane spacing ($d_{110} = 2.578$ {\AA}) is 0.44\% larger than bulk ($d_{110} = 2.567$ {\AA}) and the peak asymmetry suggests a strain gradient from smaller to larger $d_{110}$ spacings across the vertical extent of the film. Least-squares refinement from the positions of multiple off-specular reflections gives $a=2.862$ {\AA} (+0.26\%), $b=5.943$ {\AA} (+1.34\%), $c=4.988$ {\AA} (+0.67\%) and an atomic cell volume of $V=21.207$ {\AA}$^3$ (+2.3\%), where all percentages given are relative to bulk U at room temperature from Ref.~\cite{Barrett1963CrystalTemperatures}.  

At $T_\textrm{U}=200$ $^\circ$C, the specular $\alpha$-U(110) peak is instead symmetric and close to the bulk value, with lattice parameters of $a=2.858$ {\AA} (+0.14\%), $b=5.854$ {\AA} (-0.25\%), $c=4.993$ {\AA} (+0.76\%) and an atomic volume of $V=20.885$ {\AA}$^3$ (+0.65\%) indicating that the $b$-axis strain has changed from tensile to compressive while the $c$-axis expansion persists. The Laue fringes are suggestive of high crystallinity and a sharp U-Ti interface. The periodicity of the oscillations can be used to extract the crystalline ordered volume, with the agreement between $t_{\textrm{Laue}} = 500\pm10$ {\AA} and the XRR derived thickness of $t_U=514\pm$6 {\AA} suggesting that crystalline order is maintained throughout the full thickness of the U layer. The rocking curve also adopts the distinctive two-component lineshape common to many high quality thin films \cite{Gibaud1993High-resolutionSapphire,Durand2011InterpretationDeposition}. %, where the two-component peak profile indicates the presence of two distinct correlation lengths in the system .. The broad `Lorentzian-squared' component is generally attributed to short range strain and misfit dislocation defects while the narrow, resolution-limited Gaussian component is thought to arise from the long range ordering of flat atomic planes parallel to the surface \cite{Wildes2001TheSapphire}.

\begin{figure}[t]
\centering
	\includegraphics[width=0.48\textwidth]{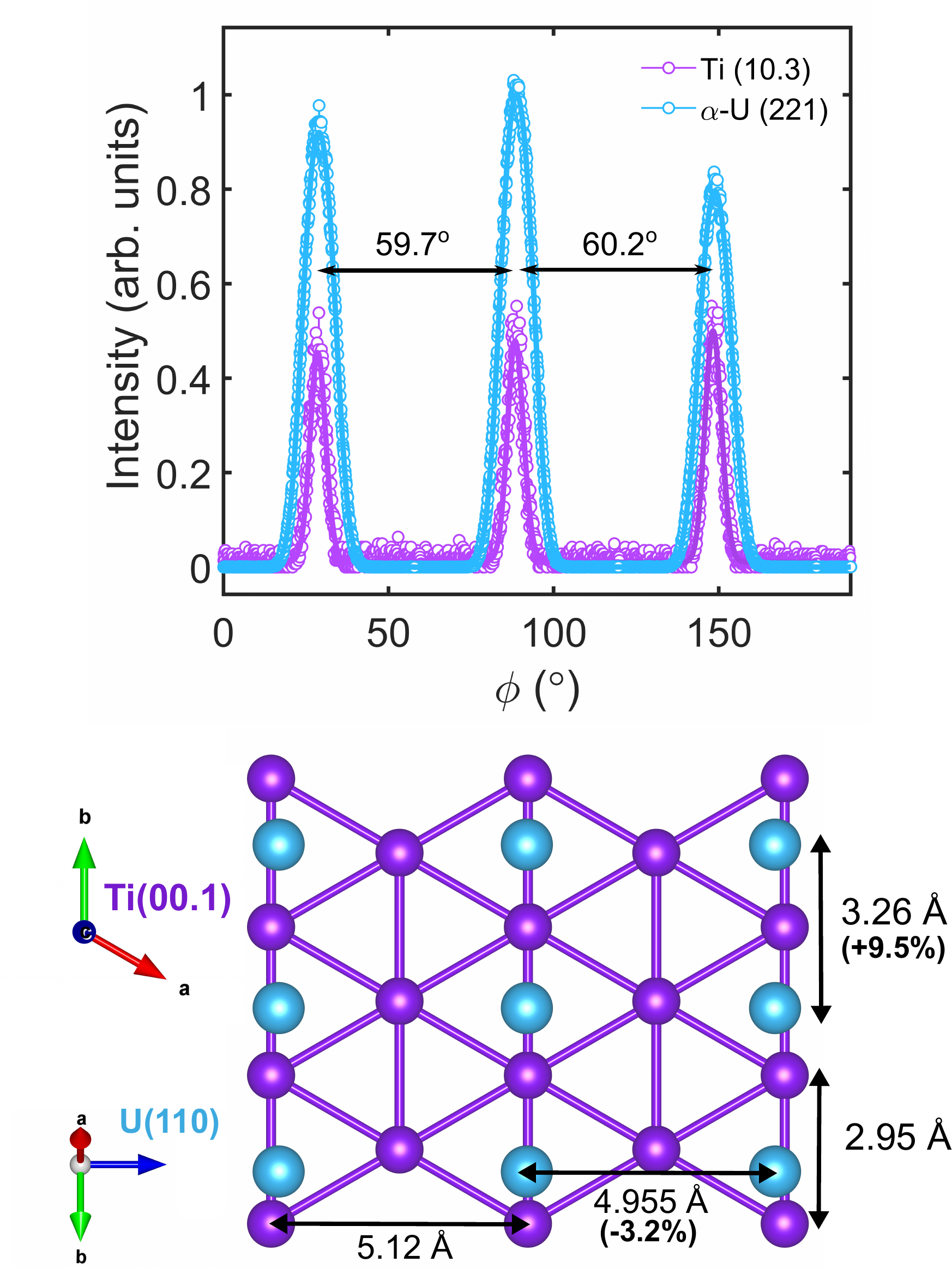}
	\caption{Top: Angular dependence of the off-specular reflections in an epitaxial Ti(00.1)/$\alpha$-U(110) sample deposited at 200$^\circ$C. Bottom: Two-dimensional illustration of the expected epitaxial match for a single U domain, showing the alignment of $d_{\mathrm{U}}=c_\textrm{U}=4.955$ {\AA} and $d_{\mathrm{Ti}}=2d_{100}=5.12$ {\AA}. All quoted lattice parameters are bulk experimental values at room temperature from literature.}
	\label{fig:UTiTotal}
\end{figure}

The in-plane epitaxial relationships in these two well-ordered systems were determined from the in-plane ($\phi$) dependence of the Ti(10.3) and U(221) off-specular reflections. The example dataset shown in the top panel of Fig.~\ref{fig:UTiTotal} indicates an approximate alignment of
\begin{align*}
     (00.1)_{\mathrm{Ti}}\parallel (110)_{\mathrm{U}} \mathrm{~ and ~} [01.0]_{\mathrm{Ti}}\parallel [1\bar{1}0]_{\mathrm{U}}.
\end{align*}
The epitaxy is likely to be governed by the match shown in the bottom panel of Fig.~\ref{fig:UTiTotal} where the misfit strain, ($d_\textrm{U}-d_\textrm{Ti})/d_\textrm{Ti}$ at room temperature is $-3.2$ \%. As hcp-Ti is six-fold symmetric in the (00.1) plane, any 60$n^\circ$ ($n\in Z$) in-plane rotation of the U layer brings the relevant planes into alignment. This should result in six energetically equivalent ways for the first monolayer of uranium to nucleate on the Ti(00.1) surface, as confirmed by the 60$^\circ$ separation of the (114)$_\textrm{U}$ peaks in the $\phi$-scan.%Given the mirror symmetries (e$.$g$.$ along the in-plane $c$ axis), this creates three distinct U domains.

The stability of epitaxial $\alpha$-U at these relatively low deposition temperatures was unexpected, as epitaxial Nb(110)/$\alpha$-U(110) and W(110)/$\alpha$-U(001) systems are typically grown at 450-600 $^\circ$C \cite{Ward2008TheUranium,Springell2008ElementalFilms,Springell2014MalleabilityFilms}. In the case of Ti/U, temperatures above 200$^\circ$C are clearly detrimental to the quality of the interface. The degradation of the XRR signal, rocking curve profile and U/Ti Laue fringes all suggest that the sharp U-Ti interface, and hence the coherent epitaxial match, has been partially lost at $T_\textrm{U}=400^\circ$C and fully lost at $T_\textrm{U}=600^\circ$C. The additional, non-elemental diffraction peaks seen in the $2\theta$-$\omega$ scans are likely to originate from Ti-rich alloys ($2\theta=36-38^\circ$) and U$_2$Ti ($2\theta_{110}=37.2^\circ$) \cite{Knapton1954TheTiU2}.

\subsection{Zirconium buffered system}

\begin{figure}[t]
\centering
	\includegraphics[width=0.45\textwidth]{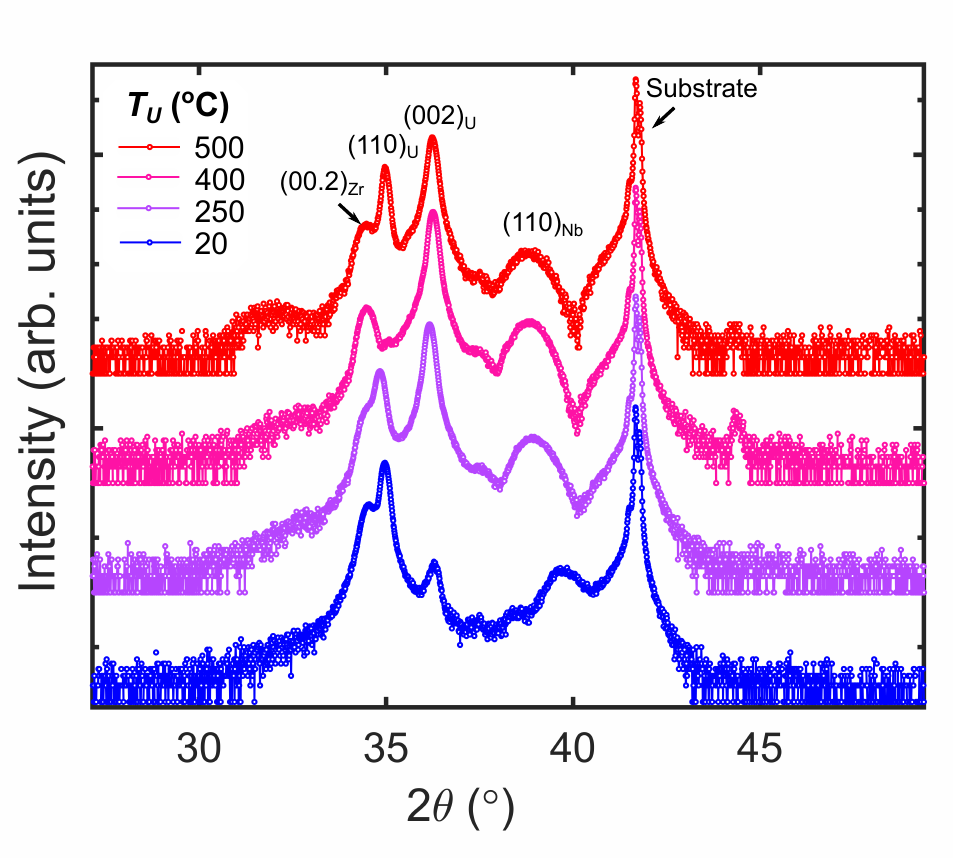}
	\caption{$2\theta$-$\omega$ scans of U layers deposited onto Zr(00.1) buffer layers at various temperatures. Elemental U and Zr peaks are present across the entire temperature range, but the dominant U orientation varies with deposition temperature. Data are offset vertically and presented on a logarithmic intensity scale for clarity.}
	\label{fig:U_Zr_HAs}
\end{figure}

A similar series was grown using zirconium buffer layers deposited onto $c$-plane sapphire at 700 $^\circ$C. These single crystal Zr(00.1) layers adopt an in-plane epitaxial relationship of Al$_2$O$_3$[10.0]$\parallel$Zr[10.0] and exhibit rocking curves with widths of 1-2$^\circ$. This limits the mosaic spread and grain size of subsequent U layers, but the epitaxial relationships are still of interest. Fig.~\ref{fig:U_Zr_HAs} shows the coupled $2\theta$-$\omega$ scans for the series.%, where elemental U and Zr peaks are present across the entire temperature range and the dominant U orientation varies with growth temperature. %

At room temperature, the spectrum is primarily $\alpha$-U(110) with small inclusions of $\alpha$-U(001). The off-specular reflections, included as Supplemental Information \cite{SupplementalInformation}, suggest an orientation relationship of 
\begin{align*}
     (00.1)_{\mathrm{Zr}}\parallel (110)_{\mathrm{U}} \mathrm{~ and ~} [10.0]_{\mathrm{Zr}}\parallel [1\bar{1}0]_{\mathrm{U}}    
\end{align*}
where the alignment of $d_{\mathrm{U}}=\frac{1}{2}(a_\textrm{U}^2+b_\textrm{U}^2)=3.26$ {\AA} and $a_{\mathrm{Zr}}=3.24$ {\AA} produces a low misfit strain of -0.6 \%. As with the Ti/$\alpha$-U(110) system, the hexagonal symmetry of the buffer facilitates six equivalent matches in 360$^\circ$. The measured $\alpha$-U lattice parameters are $a=2.853$ {\AA} (-0.015\%), $b=5.826$ {\AA} (-0.73\%) and $c=5.136$ {\AA} (+3.6\%) and the atomic volume of $V=21.346$ {\AA}$^3$ (+3.1\%) is unusually large, even for a structure as malleable as $\alpha$-U. 

As the deposition temperature is increased toward 400$^\circ$C, the sample gradually becomes pure $\alpha$-U(001), with the (002)$_\textrm{U}$ reflection gaining relatively weak Laue fringes. The orientation relationship between the layers determined from the Zr(10.5) and $\alpha$-U(023) reflections (Fig.~S3 in Supplemental Information \cite{SupplementalInformation}) is
\begin{align*}
     (00.1)_{\mathrm{Zr}}\parallel (001)_{\mathrm{U}} \mathrm{~ and ~} [10.0]_{\mathrm{Zr}}\parallel [100]_{\mathrm{U}}    
\end{align*}
where the alignment of $d_{\mathrm{U}}=5.869$ {\AA} and $d_{\mathrm{Zr}}=5.61$ {\AA} results in a large misfit strain of +4.6 \%. The refined lattice parameters are $a=2.865$ {\AA} (+0.39 \%), $b=5.827$ {\AA} (-0.71 \%), $c=4.952$ {\AA} (-0.05 \%) and the atomic volume is bulk-like at $V=20.67$ {\AA}$^3$. Again, six-fold symmetry is seen in the $\alpha$-U(023) $\phi$-scan due to the six equivalent matches with the hexagonal Zr(00.1) surface. 

The transition from an $\alpha$-U(110) layer with a low strain epitaxial match and a large atomic volume, to an $\alpha$-U(001) layer with a large misfit strain and a bulk-like atomic volume suggests it is energetically favourable for the $\alpha$-U structure to revert to a bulk-like atomic volume at the expense of the epitaxial match and quality of the interface. The formation of a interfacial U-Zr layer that may influence the epitaxy is also suggested by the data. 

A 1-2 nm reduction in $t_\textrm{U}$ with increasing $T_{\mathrm{U}}$ is seen via the XRR-derived U layer thicknesses and, at 500 $^\circ$C, the reflectivity profile no longer shows Kiessig fringes. A gradual reduction in the intensity of the (00.2)$_\textrm{Zr}$ reflection with increasing $T_{\mathrm{U}}$ also suggests the formation of a interfacial U-Zr layer that increases in thickness with $T_\textrm{U}$. The strains generated by the unusually large mismatch between the Zr(00.1) and $\alpha$-U(001) layer (+4.6 \%) may be relieved by such a transition region, facilitating the observed change in orientation and reduction in atomic volume.  %As XRR signal depends on the presence of sharp interfaces and abrupt changes in electronic density, their disappearance suggests a loss of a well-defined U-Zr interface at higher deposition temperatures. 

\begin{figure}[t]
\centering
	\includegraphics[width=0.45\textwidth]{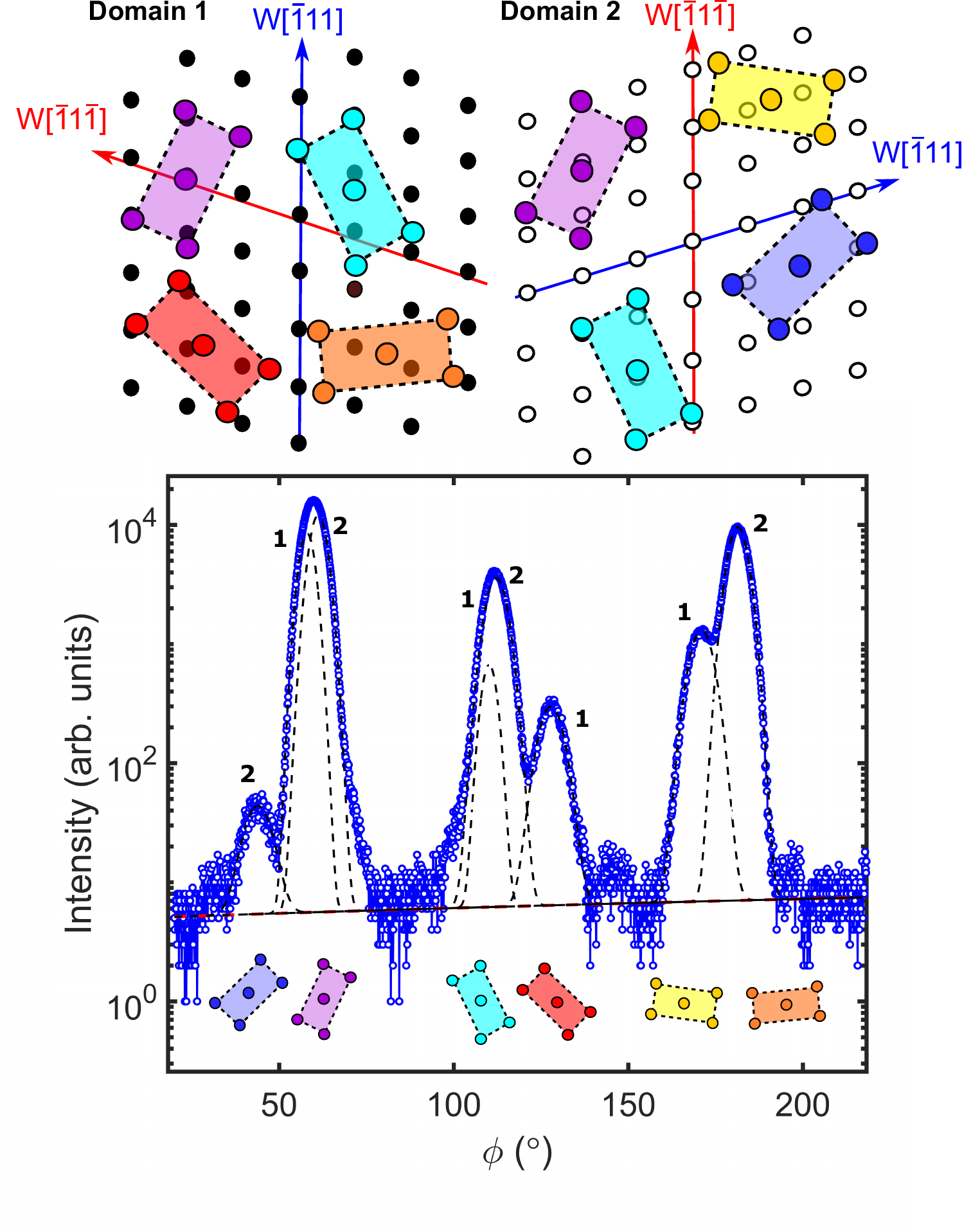}
	\caption{Top: Expected epitaxial matches for $\alpha$-U(001) deposited onto a twinned W(110) buffer layer, adapted from \cite{Ward2008TheUranium}. Tungsten domains 1 and 2 represented by filled and open circles. Colors used to represent `unique' uranium domains. Each vertical tungsten axis is aligned with Al$_2$O$_3$[00.1]. Bottom: New $\phi$-scan of the off-specular U(023) reflection in a new $\alpha$-U(001) system. Each peak has been matched to the relevant colored domains, (colored images below) and individual fitted peak components (black dotted lines) are ascribed to the relevant W domain, with labels 1 and 2, respectively.}
	\label{fig:SN1931_U023}
\end{figure}

\subsection{Tungsten buffered systems}

The growth of complex, multi-domain $\alpha$-U(001) was first reported in 2008 \cite{Ward2008TheUranium}. Ward \textit{et al.} proposed a model wherein eight domains of $\alpha$-U nucleate on a twinned W(110) buffer as a result of a close match between the distances $d_U=2.556$ {\AA} ($d_{110}$) and $d_W=2.584$ {\AA} ($2d_{112}$). The two W domains and eight U domains are illustrated in the top panel of Fig.~\ref{fig:SN1931_U023}. In this idealised system, certain U domains (shown here in light blue and purple) are `degenerate' with respect to the buffer and so a total of six peaks should be resolvable using a point detector and in-plane $\phi$-scans. However, only four domains were seen in the original study \cite{Ward2008TheUranium}. 

The bottom panel of Fig.~\ref{fig:SN1931_U023} shows a $\phi$-scan of the $\alpha$-U(023) reflections in a new, high quality W/U sample. This scan maps the relative orientations of any (010) planes (i.e. the $b$-axes) in the $\alpha$-U(001) layer. The scan shows six well-resolved peaks at angular separations that correlate well with the matches predicted by Ward \textit{et al.}. A total of \textit{eight} peaks are required for an accurate fit as there is a slight misorientation between the reflections from the two `degenerate' (light blue and purple) pairs of domains, presumably due to strains in the buffer. These strains are also likely to be the cause of the unequal peak intensities, which imply there are strong preferences for certain U orientations. The crystalline quality of the U layer is significantly improved by reducing the deposition temperature to $450$ $^\circ$C used in Ref.~\cite{Springell2014MalleabilityFilms} as opposed to 600 $^\circ$C used in Ref.~\cite{Ward2008TheUranium}.

\subsection{Niobium buffered systems}

\begin{figure*}[t!]
\centering
	\includegraphics[width=\linewidth]{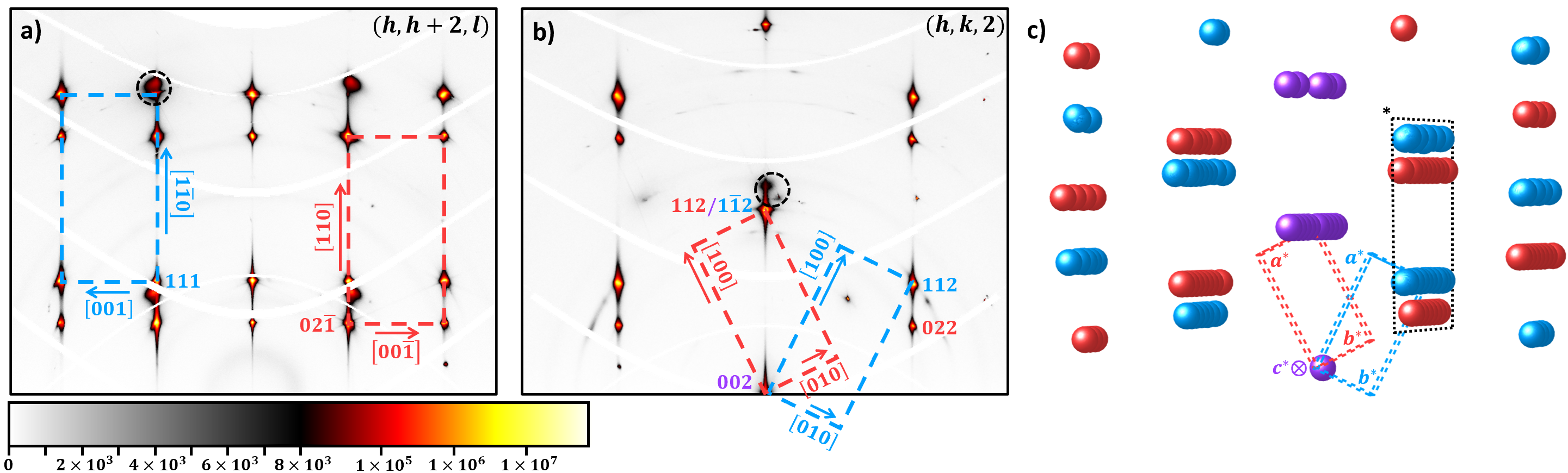}
	\caption{RSMs for a) the $(h,h+2,l)$ and b) the $(h,k,2)$ type planes reconstructed and denoted within he primary domain setting. Scale is linear white-black and $\mbox{log}_{10}$ black-red-white. Single reciprocal space net units for the primary (red) and secondary (blue) domains are shown by dashed lines with select reflections indexed. Examples of sharp sapphire reflections and weak, broad Nb buffer reflections are highlighted within black dashed circles. c) 3D reciprocal space schematic showing a selection of indexed reflections. Sphere colour corresponds to domain origin consistent with left-hand panels. Degenerate reflections in the ``specular plane'' indicated by purple spheres. Reciprocal unit cells are shown as dashed cuboids with reciprocal lattice vectors marked. A section of the $(h,h+2,l)$ type plane shown in panel (a) is highlighted by a dashed black box and asterisk.}
   \label{fig:U_Nb_ID28}
\end{figure*}

To date, all epitaxial films of $\alpha$-U(110) on Nb(110) have been reported as single domain systems where the epitaxy is governed by a uni-directional in-plane match between $d_\textrm{U}=\frac{1}{2}(a_\textrm{U}^2+b_\textrm{U}^2)^\frac{1}{2}=3.264$ {\AA} and $a_\textrm{Nb}=3.311$ {\AA} \cite{Ward2008TheUranium, Springell2008ElementalFilms}. New reciprocal space maps (RSMs) taken at the ID28 beamline (ESRF) reveal a second domain consistently missed by point-detector measurements. These domains are referred to as primary `red' and secondary `blue' in the following discussions.

This unusual situation has arisen as the ($hhl$) reflections - i.e. those commonly used to check the symmetry of the U layer in laboratory $\phi$-scans - are coincident, but the degeneracy is clearly lifted outside of this plane. The ($h$,$h$+2,$l$) RSM in Fig.~\ref{fig:U_Nb_ID28}(a) shows an example of a fully non-degenerate plane of reflections in a Nb(110)/U system, while the ($h$,$k$,2) type RSM in Fig.~\ref{fig:U_Nb_ID28}(b) demonstrates both the coincidence of reflections in the ($hhl$) plane and splitting away from this plane. Fig.~\ref{fig:U_Nb_ID28}(c) shows the location in reciprocal space for a selection of the observed reflections, where the overlapping reflections from the two distinct domains (red and blue) are represented by purple spheres. The two $\alpha$-U domains are related by an approximately $52\degree$ clockwise rotation about a shared $\bm{c}^{*}$ axis set into the page, where the experimentally determined reciprocal space transformation matrix

%Within these reconstructions, the conventional reflections (red) lie at integer Miller indices whereas the second set of equally intense uranium reflections, indicating equal domain occupation, lie at fractional Miller indices in the reference frame of the conventional domain. Introducing a secondary domain (blue) related to the primary (red) domain in the manner shown in Fig.~\ref{fig:U_Nb_ID28}(c) allows all observed reflections to be indexed successfully. This equates to an approximate $52\degree$ clockwise rotation about a shared $\bm{c}^{*}$ axis set into the page and can be expressed via the experimentally determined reciprocal space transformation matrix

\begin{equation}\label{matrixequation}
\frac{1}{4010}\begin{bmatrix}2451 & 1549 & 0\\-6461 & 2461 & 0\\0 & 0 & 4010\\\end{bmatrix}\cdot\begin{pmatrix} h_{1} \\ k_{1} \\ l_{1} \end{pmatrix} = \begin{pmatrix} h_{2} \\ k_{2} \\ l_{2} \end{pmatrix}
\end{equation}

\noindent transforms from primary to secondary Miller indices. The reverse operation is found by taking the inverse of the $3\times 3$ matrix in Eq. \ref{matrixequation}. 

\begin{figure}[b]
\centering
	\includegraphics[width=\linewidth]{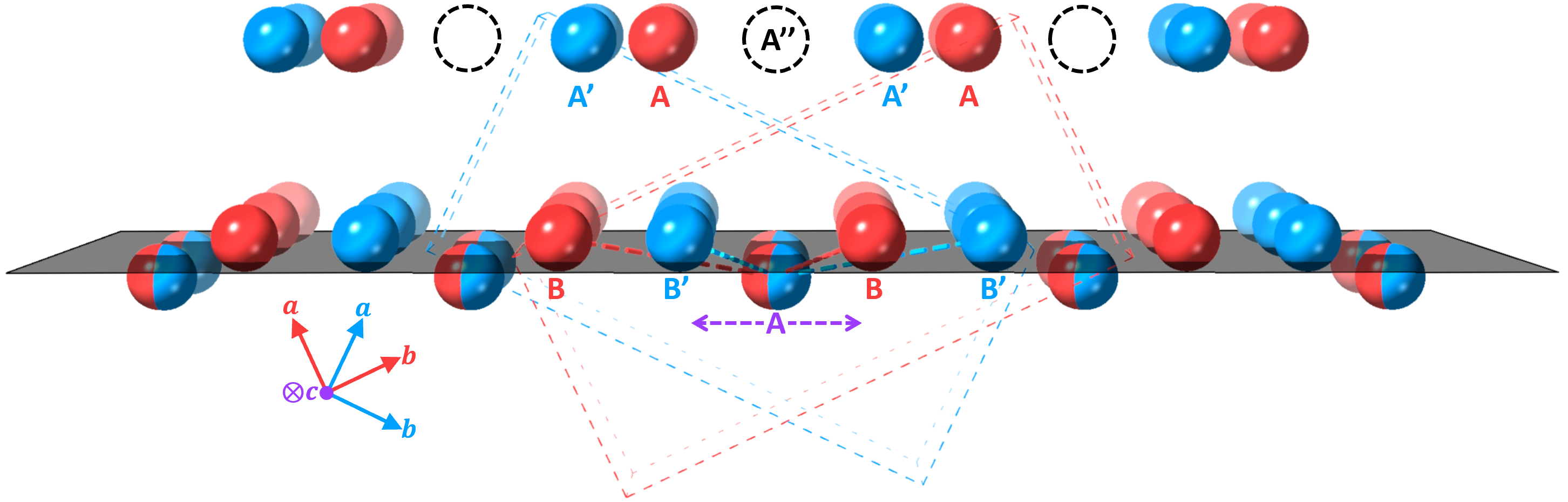}
	\caption{Schematic representation of the crystallographic relationship between the two domains. Atoms are shown as coloured spheres, unit cells by dashed cuboids and axes by compass (bottom left). Colours are consistent with Fig.~\ref{fig:U_Nb_ID28}. The growth plane (translucent grey) corresponds to the $(110)$ (primary) and $(1\bar{1}0)$ (secondary) planes. The first, degenerate, monolayer is marked with a purple label. The choice of atomic positions for the second monolayer are indicated by B/B', with example bonds shown by dashed lines, and the resulting third monolayer positions by A/A'. The hypothetical atomic positions corresponding to non-rotational domain switch are shown as dashed black circles and labelled A''.}
\label{fig:U_Nb_ID28_realspace}
\end{figure}

The origin of this `hidden' domain can be understood as follows. As U atoms are deposited onto the Nb(110) surface, each nucleation event initiates the growth of a proto-domain of $\alpha$-U with a $\langle110\rangle_\textrm{U}$ growth axis. In each of these newly forming proto-domains, the atomic arrangements in the growth plane (i.e. the lowest layer in Fig.~\ref{fig:U_Nb_ID28_realspace}) can be considered identical. However, atoms in the subsequent monolayer have an energetically degenerate choice of bonding with the long bond on the left (and short bond on the right) or the reverse, creating either a left-skewed (B) or right-skewed (B') layer. This choice fully constrains the growth axis and defines the domain. It is important to note that a $180\degree$ in-plane rotation fails to map one domain onto the other, instead stacking A/A' directly above B/B'. The secondary domain origin must be shifted in-plane to ensure that the atomic sites are coincident in layer A.

If the growth mode is purely island-like, the presence of these left- and right-skewed domains is likely to result in a columnar domain structure with in-plane anti-phase domain boundaries. Pure layer-by-layer growth would preferentially create a layer with either left or right-skewedness, as vertical switching of the `skewedness' would require the energetically unfavourable stacking of atoms almost directly above each other as shown in Fig.~\ref{fig:U_Nb_ID28_realspace}. The equal intensities of the two sets of reflections indicates equal domain occupations and the unlikelihood of direct atomic stacking suggests an island-like or mixed-type growth mode, but a fully conclusive determination of the atomic stacking pattern across the domain boundaries requires the application of a non-averaged technique, e$.$g$.$ atomic resolution transmission electron microscopy. %Note that a $180\degree$ in-plane rotation fails to map one domain onto the other as the unit cell centres are non-coincident along $\bm{c}$ and instead produces a similarly unfavourable situation stacking atoms directly above each other. 

It is important to acknowledge that this `hidden' domain should also be present in the Ti(00.1)/$\alpha$-U(110) and Zr(00.1)/$\alpha$-U(110) systems from previous sections. Indeed, these twin domains have been observed in (241) $\phi$-scans for both systems, but the scans have been omitted from this report due to their complexity. The true number of domains is then double the value suggested by the symmetry of the ($221$) $\phi$-scans.

\section{Conclusions}
Several epitaxial $\alpha$-U systems have been stabilised using both new (Ti, Zr) and known (Nb, W) elemental metallic buffer layers, with some U layers forming well-ordered systems without the need for substrate heating. The range of epitaxial strains and atomic volumes observed are expected to produce significant variations in the low temperature electronic properties. A combination of magnetotransport (e$.$g$.$ Hall coefficient, resistivity and magnetoresistance) and synchrotron diffraction measurements can now be used to probe the interplay between the CDW and SC ground states in these $\alpha$-U thin films. Specifically, studies of the superconducting transitions in Ti, Zr and W buffered samples (where $T_\textrm{c}< 1$ K in the buffer) should be used to probe the long-standing issue of unpredictable superconductivity in bulk $\alpha$-U crystals. Measurements of superconductivity in the Nb buffered systems will be more challenging ($T_\textrm{c}\approx9.2$ K in bulk Nb), but $T_\textrm{c}$ in the buffer could be suppressed by reducing its thickness or adding magnetic impurities. The complex structural information determined here will be essential for the accurate analysis and understanding of future electronic transport measurements conducted using $\alpha$-U thin films.

\begin{acknowledgments}
This research was supported by the Engineering and Physical Sciences Research Council (EPSRC), UK, through the Centre for Doctoral Training in Condensed Matter Physics (CDT-CMP) grant no. \textsc{EP/L015544/1}, and the National Nuclear User Facility for Radioactive Materials Surfaces (NNUF-FaRMS), grant no. \textsc{EP/V035495/1}, \cite{farms}. We also acknowledge the European Synchrotron Radiation Facility (ESRF) for provision of synchrotron radiation facilities. 
\end{acknowledgments}

\end{document}